\documentstyle[preprint,epsfig,prb,aps]{revtex} 

\def\mib#1{_{\rm#1}}
\def\mit#1{^{\rm #1}}

\def\vecG{{\bf G}}
\def\vecr{{\bf r}}
\def\veck{{\bf k}}
\def\vecq{{\bf q}}

\begin{document}

\draft

\title{Time-dependent screening of a positive charge distribution in metals: Excitons on an ultra-short time 
scale}

\author{Wolf-Dieter Sch{\"o}ne and Walter Ekardt}
\address{Fritz-Haber-Institut der Max-Planck-Gesellschaft, 
         Faradayweg 4-6, 14195 Berlin, Germany}
       
\date{\today}

\maketitle

\begin{abstract}
Experiments determining the lifetime of excited electrons in crystalline copper reveal states which cannot be 
interpreted as Bloch states [S. Ogawa {\it et al.\/}, Phys. Rev. B {\bf 55}, 10869 (1997)]. In this article 
we propose a model which explains these states as transient excitonic states in metals. The physical 
background of transient excitons is the finite time a system needs to react to an external perturbation, in 
other words, the time which is needed to build up a polarization cloud. This process can be probed with 
modern ultra-short laser pulses. We calculate the time-dependent density-response function within the jellium 
model and for real Cu. From this knowledge it is possible within linear response theory to calculate the time 
needed to screen a positive charge distribution and -- on top of this -- to determine excitonic binding 
energies. Our results lead to the interpretation of the experimentally detected states as transient excitonic 
states. 
\end{abstract}



\section{Introduction}
\label{Intro}

\begin{figure} 
\begin{center}
\epsfxsize=15.0cm
\centerline{\epsffile{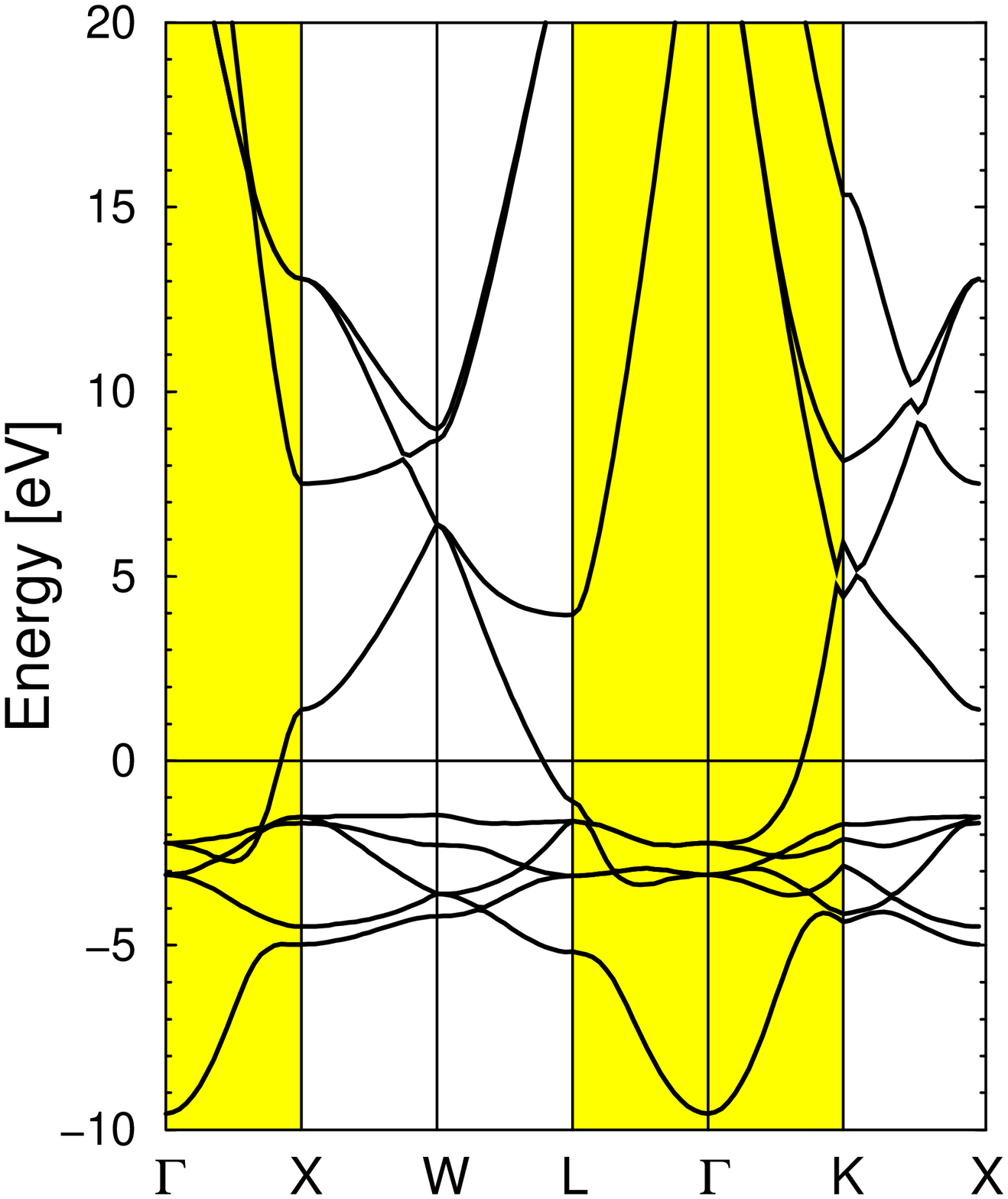}}
\end{center}

\caption{
The band structure of Cu calculated using a plane-wave code. \cite{fhi96} For the energy range between 1 and 
4 eV above $E_F$ Cu possesses no band states in the $\Gamma-L$ and the $\Gamma-X$ directions. There is a {\it 
sp\/}-like band in the $\Gamma-K$ direction. 
} 
\label{Cubands} 
\end{figure}

Measurements of the lifetime of excited electrons in Cu \cite{PetekCu,HertelCu} using time-resolved
two-photon photoemission (TR-2PPE) spectroscopy \cite{PetekRev} in connection with very short laser pulses
reveal a couple of amazing results.  Most stunning is the fact that the data show lifetimes of states along
the (100) and (111) direction ($\Gamma-X$ and $\Gamma-L$, respectively) with energies between 1 and 4 eV
above the Fermi energy $E_F$.  As can be seen from the band structure shown in Fig.~\ref{Cubands}, Cu has
no Bloch states at these energies in these two directions.  But also the data for the lifetime of electrons
in the remaining direction (110) -- the $\Gamma-K$ direction -- are puzzling.  {\it Ab initio\/}
calculations of the lifetime of hot electrons along the (110) direction using the $GW$ approximation
\cite{wdCu,wdNoble} reveal a good agreement with the experimental data for electron energies between 2 and
3 eV.  For energies lower than 2 eV the results of the $GW$ calculation deviate markedly from the
experimental data.  This is particularly remarkable because typically many-body calculations within the
$GW$ approximation agree very well with experimental data.  This holds for both, the real part of the
resulting quasiparticle energies \cite{louie,GWrev} as well as its imaginary part,
\cite{wdCu,wdNoble,igorlife,igorMeLife} respectively.  The imaginary part of the quasiparticle energy of a band state
is inverse proportional to the lifetime of this state.  \cite{mahan} Therefore the discrepancy of the
theoretical and experimental results must be due to physics which is not covered by the $GW$ approximation.
The last point about the experimental data we would like to mention is the fact that the lifetimes in the
three directions are very similar \cite{PetekCu} although the band structure gives no hint for this 
experimental finding.

The discussion in the preceding paragraph shows that TR-2PPE experiments detect states which are 
obviously no band states (which are single-particle states). Consequently there has been some 
speculation in the literature about the origin of these states, most of which violate well 
established fundamental physical laws. The only explanation which is consistent in this respect and 
does involve only single-particle states is that electrons which originally had a momentum in the, e.g.,  
(110) direction are scattered from a (110) low-energy electron diffraction (LEED) state into a, 
say, (100) LEED state. This would explain why experiments measure the lifetime in this direction. 
However, the cross section for this process should be small. Moreover, this mechanism would not explain 
the difference between the results of calculated lifetimes within the $GW$ approximation and 
the experimental ones. In order to escape these difficulties Cao {\it et al.\/} speculated 
\cite{CaoCu} that ``the strongly localized 3{\it d\/} holes generated by photoexcitation of 3{\it 
d\/} electrons can trap excited electrons through attractive Coulomb interaction''. We picked up 
this suggestion and outlined the physical mechanism of excitons in metals on a very short time 
scale, so-called transient excitons. \cite{WEexc} In this article we present a model of how to 
actually calculate these states.

Now it is common wisdom that excitons exist in semiconductors and rare gases but not in metals.
\cite{knox} However, in these cases one is thinking of stationary conditions under which the valence
electrons of the system have enough time to built up a polarization cloud.  For large times the initial
Coulomb interaction between the hole and the electron transforms into a Thomas-Fermi-like potential in
metals \cite{PiNo} and a Coulomb potential which is screened by the macroscopic dielectric constant in
semiconductors \cite{AshMer}.  But this polarization needs a final time to be built up and it is during
this time that transient excitons can exist.  \cite{WEexc} Consequently transient excitons cannot be
detected with traditional photoabsorption spectroscopy.  It needs modern TR-2PPE spectroscopy combined with
the ultra-short laser pulses which are available nowadays.

The article is organized as follows. In Sec.~\ref{theory} we present a model that allows us to 
calculate -- and therefore proof the existence of -- transient excitonic states. In the two 
following sections we apply this model to the jellium model and to real copper. We finish with some 
estimates about the lifetime of electrons in transient excitonic states. 
 
 
\section{A model for transient excitons in metals}
\label{theory}

In this section we present a model for the calculation of transient excitonic states. A complete treatment of 
the subject implies the solution of the Bethe-Salpeter equation (BSE) \cite{BSE1,BSE2,BSE3} using the 
time-dependent particle-hole potential $v\mib{eh}(t)$ as interaction. However, it is not our objective to 
perform an {\it ab initio\/} calculation in order to calculate the excitonic energies to high accuracy. The 
intent of this article is to show that transient excitonic states in metals do exist. For this, a simple but 
nevertheless realistic model is very useful.

\begin{figure} 
\epsfxsize=15.0cm
\centerline{\epsffile{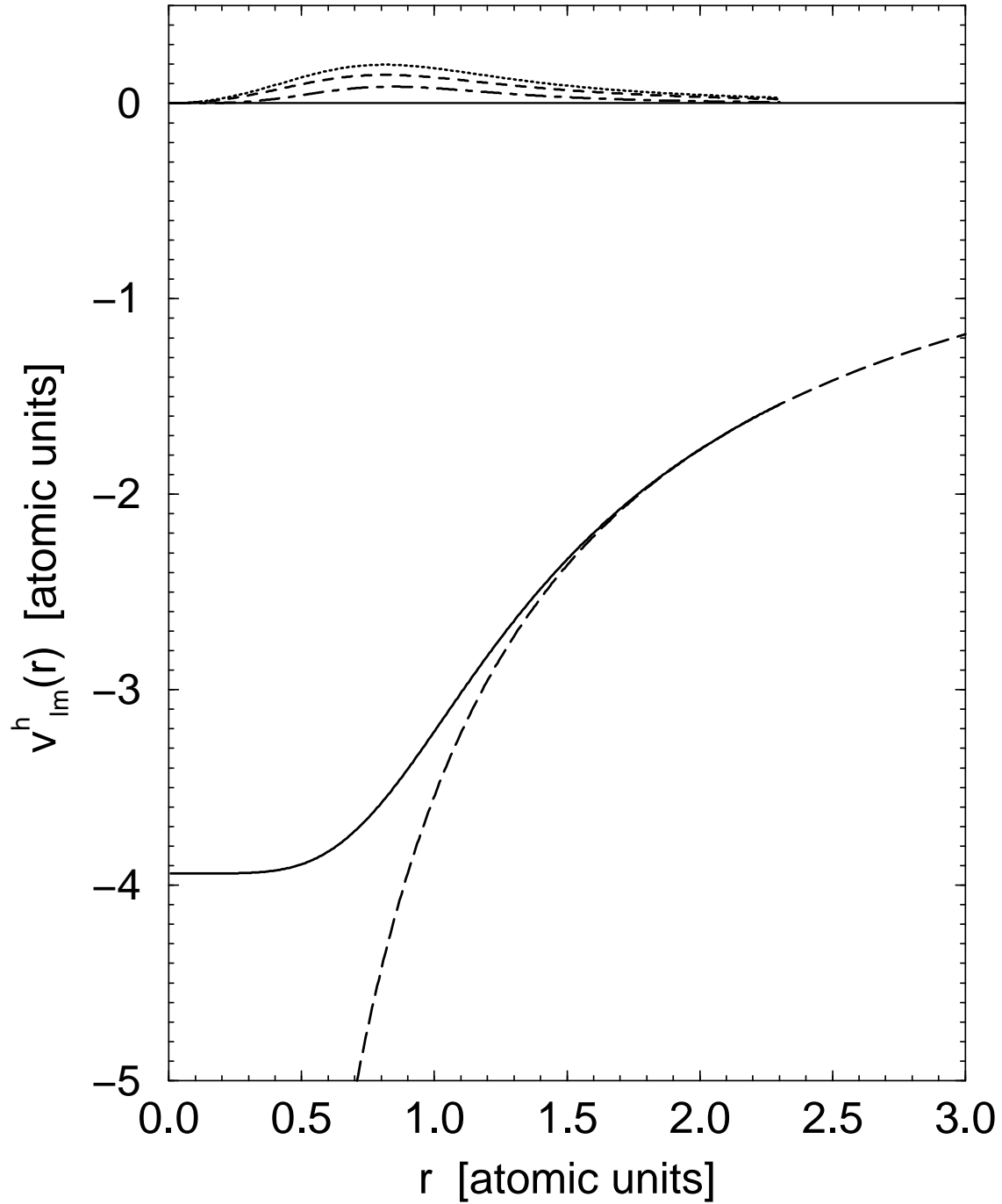}}
\caption{
The first four non-zero components of the hole potential $v^h(\vecr)$ calculated from the density 
distribution of a {\it d\/} electron at the $X$ point in Cu. The solid line is the spherical symmetric 
component (l=0, m=0). The dotted, dashed, and dot-dashed lines denote the components (l=2, m=0), (l=2, m=1), 
and (l=4, m=4), respectively. The long-dashed line shows the Coulomb potential multiplied by $\sqrt{4\pi}$.
} 
\label{holefig} 
\end{figure}

Starting point of our discussion is the potential due to a localized hole in a 3{\it d\/} band. We obtain it 
from the density of a 3{\it d\/}-band electron as
\begin{equation}
\label{dhole}
n^h(\vecr) = \cases{\alpha \, |\varphi_{3d}(\vecr)|^2   & $\vecr \in$ MT  \cr
                    0                                   &  else           \cr}    \; ,
\end{equation}
where the Bloch function $\varphi_{3d}(\vecr)$ is the solution of the Kohn-Sham (KS) equations \cite{KoSh}
\begin{equation}
\label{KSeq}
\big[ -\frac{\hbar^{2}}{2 m} {\bf \nabla}^2 \, + \,
            v\mib{eff}(\vecr) \big] \, \varphi_{\vecq,j}(\vecr) =
    \epsilon_{\vecq,j} \, \varphi_{\vecq,j}(\vecr)   \; .
\end{equation}
$\vecq$ and $j$ denote a wave vector in the Brillouin zone (BZ) and a band index, respectively.
$v\mib{eff}(\vecr)$ is the mean-field potential in which the KS electrons move. Its exchange-correlation 
potential is treated within the local density approximation (LDA). \cite{LDA} The wave functions 
$\varphi_{\vecq,j}(\vecr)$ are expanded with respect to plane waves and the interaction between the ionic 
cores and the valence electrons is described by an {\it ab initio} norm-conserving pseudopotential. 
\cite{wdCu,fhi96} $\alpha$ has to be chosen in such a way that 
\begin{equation}
\label{norm}
1 = \int_{\rm MT} d^3r \, n^h(\vecr) \; .
\end{equation}
In Cu we use for the radius of the sphere MT a value of 2.3 Bohr. The 3{\it d\/}-hole density $n^h(\vecr)$ is 
expanded with respect to spherical harmonics and using Poisson's equation the potential corresponding to 
$n^h(\vecr)$ is obtained as 
\begin{equation}
\label{vhole}
v^h(\vecr) = \sum_{l,m} \, \frac{4\pi e}{2l+1} \, \int_0^{\infty} dr' \, r'^2 \, \frac{r_{<}^{l}}{r_{>}^{l+1}} \,
             n_{l,m}^h(r') \, Y_{l,m}(\hat r)   \; .
\end{equation} 
$n_{l,m}^h(r)$ are the ($l,m$) components of $n^h(\vecr)$. $r_{<}$ and $r_{>}$ denote the smaller and larger 
length of $r$ and $r'$, respectively. $\hat r$ represents the angular part of $\vecr$. Figure \ref{holefig} 
shows the first non-zero components of $v^h(\vecr)$. The by far dominating component is the spherical 
symmetric part. In the following we approximate $v^h(\vecr)$ by its spherical part and treat this as an 
external potential switched on at $t=0$, 
\begin{equation}
\label{vext}
v\mib{ext}(r,t) = 4\pi e \, \int_0^{\infty} dr' \, r'^2 \, \frac{1}{r_{>}} \, n_{0,0}^h(r') \, Y_{0,0}(\hat r)
                \, \Theta(t) \; ,
\end{equation}
perturbing the system.

The response of the system to the perturbation $v\mib{ext}(r,t)$ is calculated within linear response theory 
(LRT). \cite{LRT} In LRT the induced density due to the perturbing potential is given by 
\begin{equation}
\label{nindrt}
n\mib{ind}(\vecr,t) = \int d^3r' \int_{-\infty}^{\infty} dt' \;
                      \chi(\vecr,\vecr';t-t') \; v_{\rm ext}(\vecr',t')      \;, 
\end{equation}
where $\chi(\vecr,\vecr';t-t')$ is the (retarded) density-response function. Since we are dealing with a 
system with translation symmetry it is useful to transform Eq.~(\ref{nindrt}) to Fourier space,
\begin{equation}
\label{nindqt}
n\mib{ind}(\vecq+\vecG,\omega) = 
          \sum_{\vecG'} \; \chi_{\vecG,\vecG'}(\vecq,\omega)   \; 
          v_{\rm ext}(\vecq+\vecG',\omega)          \; .
\end{equation}
$\chi_{\vecG,\vecG'}(\vecq,\omega)$ is obtained from the polarizability \cite{adolfo}
\begin{eqnarray}
\label{pol}
P_{\vecG,\vecG'}(\vecq,\omega) &=&
\frac{2}{V} \sum_{\veck}^{\rm BZ} \sum_{j,j'} \; 
\frac{f_{\veck,j} - f_{\veck+\vecq,j'}}
     {\hbar\omega + \epsilon_{\veck,j} - \epsilon_{\veck+\vecq,j'} + i \, \eta}  \nonumber \\
& & \qquad
     <\veck,j|e^{-i(\vecq+\vecG) \hat{\bf x}}|\veck+\vecq,j'> \;
     <\veck+\vecq,j'|e^{i(\vecq+\vecG') \hat{\bf x}}|\veck,j>
\end{eqnarray}
by solving the matrix equation \cite{adolfo}
\begin{eqnarray}
\label{chi}
\chi_{\vecG,\vecG'}(\vecq,\omega) 
& = & P_{\vecG,\vecG'}(\vecq,\omega) \nonumber \\
& + & \, \sum_{\vecG_1,\vecG_2} P_{\vecG,\vecG_1}(\vecq,\omega) \;
         \Biggr[ \delta_{\vecG_1,\vecG_2} \; v_{\vecG_1}(\vecq) \, + \, 
         f\mit{xc}(\vecG_1-\vecG_2) \Biggl] \; \chi_{\vecG_2,\vecG'}(\vecq,\omega)
\end{eqnarray}
for each wave vector $\vecq$ and frequency $\omega$, respectively. In Eq.~(\ref{pol}) the $f_{\veck,j}$ 
denote the occupation numbers, $\epsilon_{\veck,j}$ the eigenvalues, and  $<\vecr|\veck,j>=\varphi_{\veck,j}(\vecr)$ the wave functions as obtained from solving the KS equations 
(\ref{KSeq}). $V$ is the normalization volume. The sums run over all wave vectors $\veck$ in the BZ and the band indices $j$ and $j'$, respectively. In Eq.~(\ref{chi}) $v_{\vecG}(\vecq)$ is the Coulomb potential, $v_{\vecG}(\vecq)=4\pi e^2 / |\vecq+\vecG|^2$, and $f\mit{xc}(\vecG)$ represents a vertex correction connected to the local-field factor $G$ by $f\mit{xc}(\vecG) = -v_{\vecG}(\vecq) \, G_{\vecG}(\vecq)$. Within the LDA $f\mit{xc}(\vecG)$ depends only on the reciprocal lattice vector and not on the wave vector. Setting $f\mit{xc}(\vecG)=0$ results in the random phase approximation (RPA) for $\chi$. The time-dependent local density approximation (TDLDA) is given by $f\mit{xc}(\vecG)=\int d^3r \, e^{-i \vecG \vecr} dv\mib{xc}(\vecr)/dn(\vecr)$, where $v\mib{xc}(\vecr)$ is the exchange-correlation potential used in the KS equations (\ref{KSeq}). 

Having calculated the induced density $n_{\rm ind}(\vecq+\vecG,\omega)$, the induced potential 
$v\mib{ind}(\vecq+\vecG,\omega)$ is obtained via Poisson's equation. Adding the external potential 
$v\mib{ext}(\vecq+\vecG,\omega)$ and using the Kramers-Kronig relation, \cite{PiNo} a Fourier transform with 
respect to frequency yields the following expression for the total potential,
\begin{eqnarray}
\label{vtotqt}
v\mib{tot}(\vecq+\vecG,t) & = &\sum_{\vecG'} \bar v\mib{ext}(\vecq+\vecG') \times \nonumber \\
& & \; 
\Biggl\{ \delta_{\vecG,\vecG'} \, + \, \frac{4 \pi e^2}{|\vecq+\vecG|^2} \, \frac{1}{\pi}
         \int_{-\infty}^{\infty} d\omega \, \frac{1}{\omega} \, {\rm Im} \chi_{\vecG,\vecG'}(\vecq,\omega) \,
         \biggl( 1 -e^{-i \omega t} \biggr) \Biggr\} \, \Theta(t)     \; ,
\end{eqnarray}
where $\bar v\mib{ext}(\vecq+\vecG)$ is the frequency-independent part of $v\mib{ext}(\vecq+\vecG,\omega)$. A 
final Fourier transform leads to the total potential as a function of $\vecr$ and t,
\begin{equation}
\label{vtotrt}
v\mib{tot}(\vecr,t) = \frac{1}{V} \sum_{\vecq}^{\rm BZ} \sum_{\vecG}
                            e^{i (\vecq+\vecG) \vecr} \, v\mib{tot}(\vecq+\vecG,t) \; .
\end{equation}

 
\section{The jellium model}
\label{jellium}

Before we treat real Cu with our model, let us first apply it to the jellium model. Even here it leads to a 
couple of interesting results. In the jellium model, where there are no {\it d\/} bands, we use for the 
external potential a bare Coulomb potential
\begin{equation}
\label{vextJel}
v\mib{ext}(r,t) = -\frac{e^2}{r} \, \Theta(t) \; ,
\end{equation}
i.e., we calculate the time-dependent screening of a suddenly created point charge in an electron gas. 
\cite{canright} Because of the isotropy assumed in the jellium model the potential, polarizability, and 
density-response function are simple scalar quantities. Equation~(\ref{pol}) reduces to the Lindhard function 
\cite{PiNo} and Eq.~(\ref{chi}) becomes a scalar equation. So Eq.~(\ref{vtotqt}) simplifies to 
\begin{equation}
\label{vqtJel}
v\mib{tot}(q,t) = \frac{4 \pi e^2}{q^2} \, \Biggl\{ 1 \, + \, \frac{4 \pi e^2}{q^2} \, \frac{2}{\pi}
         \int_{0}^{\infty} d\omega \, \frac{1}{\omega} \, {\rm Im} \chi(q,\omega) \,
         \biggl( 1 - \cos(\omega t) \biggr) \Biggr\} \, \Theta(t)     \; ,
\end{equation}
where we used the fact that ${\rm Im} \chi(q,\omega)$ is an odd function with respect to $\omega$. 
Consequently Eq.~(\ref{vtotrt}) becomes
\begin{equation}
\label{vrtJel}
v\mib{tot}(r,t) = \frac{1}{(2\pi)^3} \int d^3q \, e^{i \vecq \vecr} \, v\mib{tot}(q,t) \; .
\end{equation}
In the jellium model ${\rm Im} \chi(q,\omega)$ has a delta-function-like shape for small wave vectors $q$, 
representing the plasmon excitation. It is centered at the plasmon frequency $\omega_p$ with $\omega_p^2=4\pi ne^2/m$, where $n$ is the electronic density. In order to handle this structure a numerical broadening $\eta$ has to be introduced. If not said otherwise, we use a broadening of $\eta=0.01$ eV, a value corresponding to the real 
width of the plasmon excitation in alkalines. \cite{KuAd} In order to be sure to sample the plasmon peak 
properly we used a frequency mesh with a spacing of 0.5 meV (and less) for wave vectors up to $1.5 \, k_F$. 

\begin{figure} 
\epsfxsize=15.0cm
\centerline{\epsffile{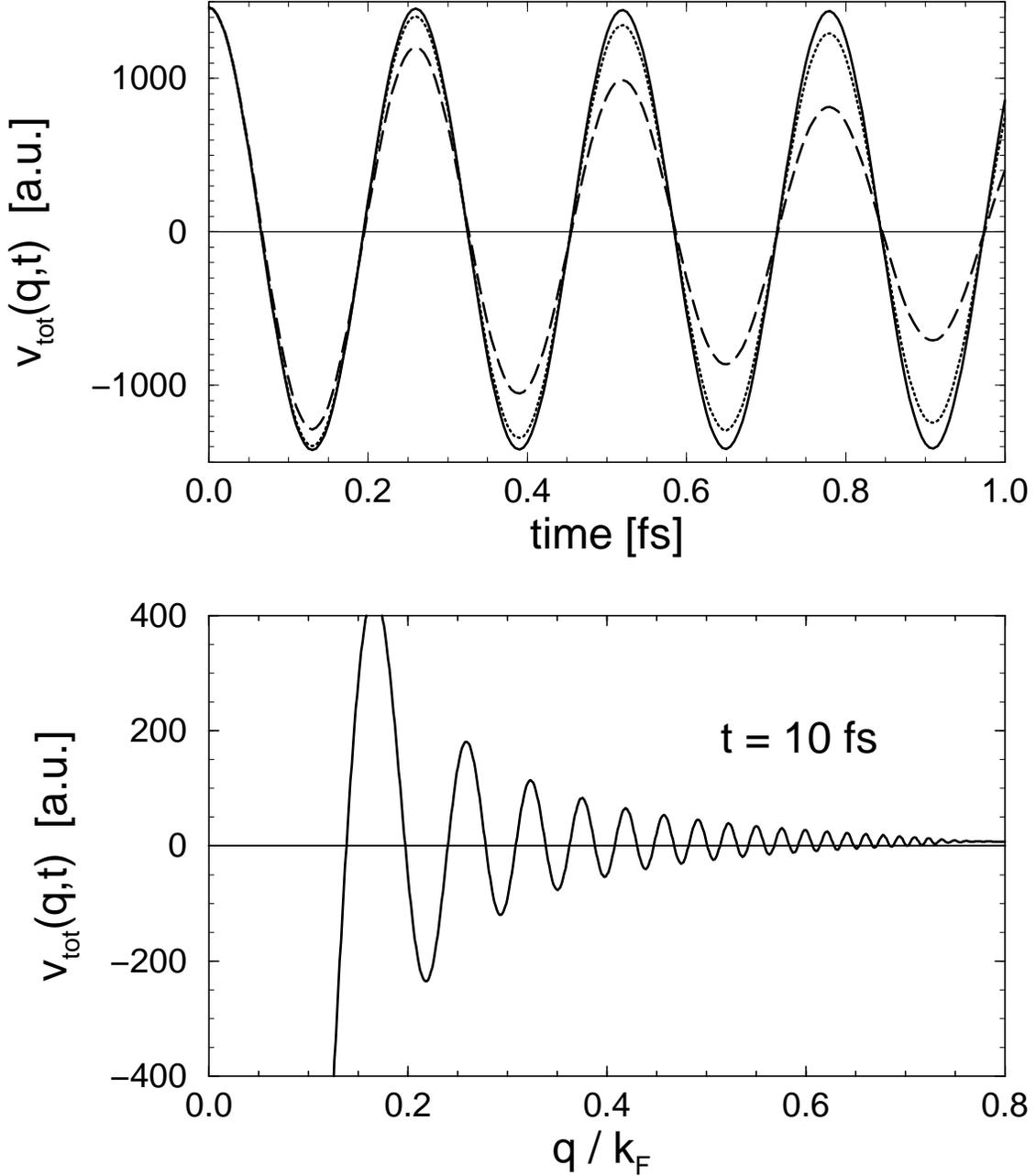}}
\caption{
The total potential $v\mib{tot}(q,t)$ for $r_s=2.07$ in atomic units. The upper panel shows $v\mib{tot}(q,t)$ for $q=0.1 \, k_F$ plotted as a function of time for different values of $\eta$ in a small time window. The broadenings used are 0.01 eV, 0.1 eV, and 0.5 eV (solid, dotted, and dashed line, respectively). The oscillations with the frequency $\omega_p$ can be clearly seen. In the lower panel $v\mib{tot}(q,t)$ is plotted as a function of $q$ in units of the Fermi wave vector for $t=10$ fs. 
} 
\label{vqt_Jel} 
\end{figure}

The upper panel of Fig.~\ref{vqt_Jel} shows a plot of $v\mib{tot}(q,t)$ as a function of $t$ for a Wigner-Seitz radius of $r_s=2.07$ and a wave vector of $q=0.1 \, k_F$, $k_F$ being the Fermi wave vector. $\chi$ was calculated within the RPA. The figure displays $v\mib{tot}(q,t)$ for three different values of $\eta$ in a time window of just one femtosecond. The solid, dotted, and dashed curve correspond to values of $\eta$ of 0.01 eV, 0.1 eV, and 0.5 eV, respectively. Even on this short time scale the influence of $\eta$ can be noticed; $v\mib{tot}(q,t)$ decays at different rates. The frequency of $v\mib{tot}(q,t)$ for the three values is nevertheless the same. It is the plasmon frequency $\omega_p$. For $r_s=2.07$ it is $\omega_p=15.83$ eV which corresponds to a period of 0.26 fs. That the potential for small wave vectors oscillates with the plasmon frequency can also be seen if ${\rm Im}\chi(q,\omega)$ is approximated by the plasmon-pole approximation (PPA)
\begin{equation}
\label{ppa}
\lim_{q\to0} {\rm Im} \chi(q,\omega) =
-\frac{\pi}{2v_q} \, \omega_p \, \Bigl\{ \delta(\omega-\omega_p) -
                                      \delta(\omega+\omega_p) \Bigr\} \, ,
\end{equation}
where $v_q = 4\pi e^2/q^2$. In this case the frequency integral in Eq.~(\ref{vqtJel}) can be done 
analytically leading to
\begin{equation}
\label{vqtPPA}
v\mib{tot}\mit{PPA}(q,t) = \frac{4 \pi e^2}{q^2} \, \Theta(t) \, \cos(\omega_p t)  \, .
\end{equation}

In the case of small wave vectors the total potential should approach the Thomas-Fermi potential \cite{mahan}
\begin{equation}
\label{TFpot}
v\mit{TF}(q) = \frac{4 \pi e^2}{q^2+q\mib{TF}^2}       \, ,
\end{equation}
for large times. $q\mib{TF}$ is the Thomas-Fermi wave vector, $q\mib{TF}^2=6\pi n e^2 / E_F$. Note that 
Eq.~(\ref{TFpot}) is derived for the limiting case of $q\to 0$ and $\omega \to 0$. Therefore it cannot be 
expected that $\lim_{t \to \infty} v\mib{tot}(q,t) = v\mit{TF}(q)$ holds for large wave vectors. For small wave vectors it is fulfilled to a high degree. As an example consider $v\mib{tot}(q,t)$ for $q=0.1 \, k_F$ and $t \to \infty$. For $\eta=0.01$ eV the difference to the Thomas-Fermi value of 10.57 a.u.\ is just 0.8\%. The deviation increases for larger broadenings. 

Having discussed the total potential as a function of wave vector we now turn to $v\mib{tot}(r,t)$. In order 
to perform a good quadrature of the integral in Eq.~(\ref{vrtJel}) we calculated $v\mib{tot}(q,t)$ on a wave 
vector mesh which is dense enough to sample the oscillations of $v\mib{tot}(q,t)$ with respect to $q$. They 
are shown in the lower panel of Fig.~\ref{vqt_Jel} where we plot $v\mib{tot}(q,t)$ as a function of wave vector at an intermediate time. In general the frequency of the oscillations increases with increasing $t$ and with 
increasing $q$. At the same time the amplitude of the oscillations die out quicker so that at large times the 
asymptotic potential is approached. The plot shows another feature which holds for all times; the oscillations stop between 0.7 and 0.8 $k_F$, i.e., they stop in the wave vector range where the plasmon excitation is destroyed by particle-hole transitions (Landau damping) and the density-response function becomes dominated by single-particle transitions. 

\begin{figure} 
\epsfxsize=12.0cm
\centerline{\epsffile{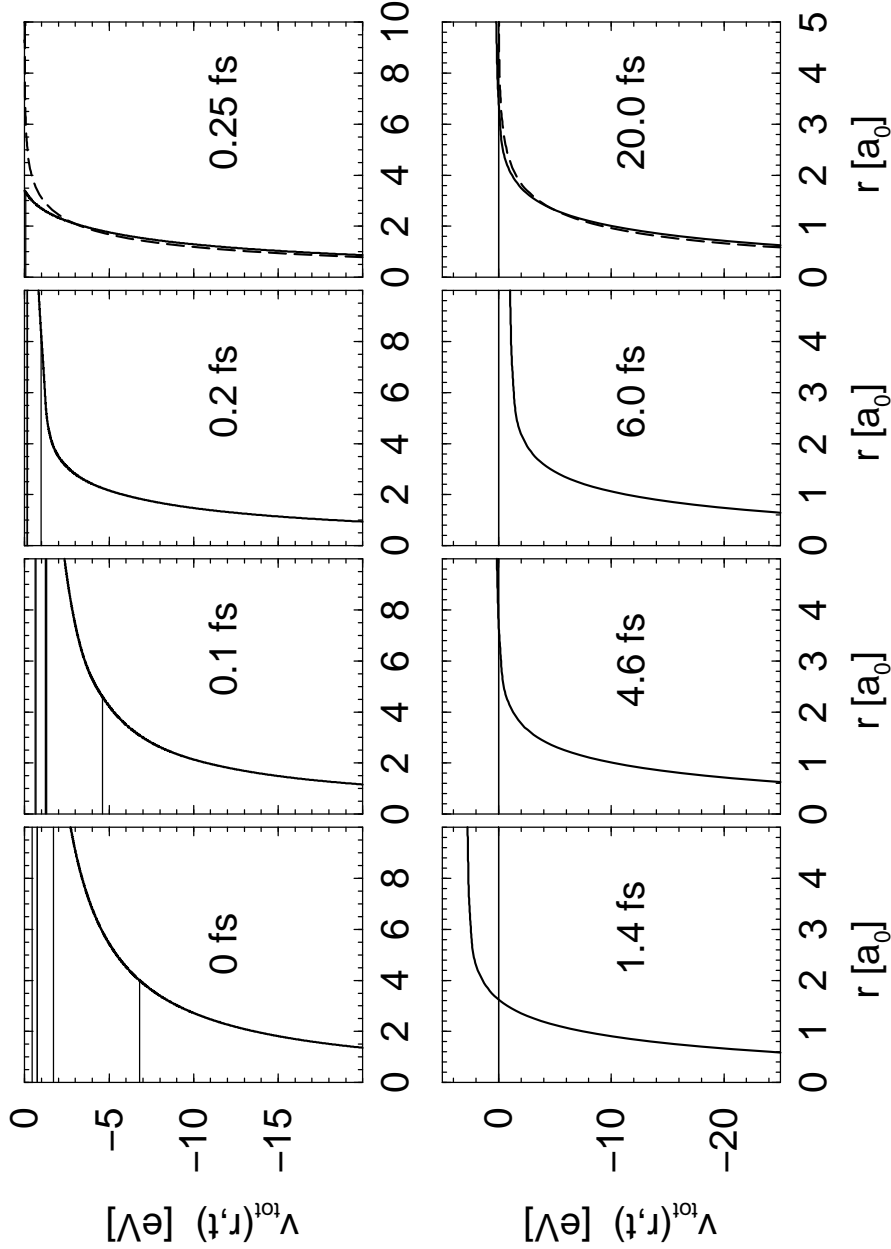}}
\caption{
Snapshots of the total potential $v\mib{tot}(r,t)$ as a function of $r$ for different times. The upper row shows the potential for small times calculated using $r_s=5$. The horizontal lines denote the eigenvalues of the potential calculated with a reduced mass of $\mu=0.5$. In the lower row $v\mib{tot}(r,t)$ is shown for larger times and $r_s=2.07$. It can be nicely seen how the system overscreens the perturbing potential which results in a decaying oscillation around the asymptotic form of the total potential. Note that the times shown cannot be used as a measure for the frequency of the oscillations which is much larger (see Fig.~\ref{vqt_Jel}). The dashed lines denote the Thomas-Fermi potentials for the two Wigner-Seitz radii.
} 
\label{vrt_Jel} 
\end{figure}

Figure \ref{vrt_Jel} shows snapshots of $v\mib{tot}(r,t)$ as a function of $r$ for fixed times calculated 
for $r_s=5$ (upper row) and $r_s=2.07$ (lower row). At $t=0$, $v\mib{tot}(r,t)$ is just the Coulomb potential (upper left plot). The horizontal lines are the first four eigenvalues for this potential calculated using a reduced mass of $\mu=0.5$ due to the equal masses of the hole and the electron, respectively. The next plot shows $v\mib{tot}(r,t)$ at a larger time. The potential is already narrower and consequently the eigenvalues are shifted upwards. This trend continues and at $t=0.25$ fs $v\mib{tot}(r,t)$ crosses the Thomas-Fermi potential [Fourier transform of Eq.~(\ref{TFpot})]
\begin{equation}
\label{vTFrt}
v\mit{TF}(r) = -\frac{e^2}{r} \, e^{-q\mib{TF}\, r}       \, ,
\end{equation}
(dashed line) for the first time. At this time the potential has no bound states any more. The system is now overscreening the perturbation which leads to a decaying oscillation of $v\mib{tot}(r,t)$ around its asymptotic form which lasts for 10 to 20 fs, the exact value depending on $r_s$. This is shown for $r_s=2.07$ in the lower panel of Fig.~\ref{vrt_Jel}. At $t=20$ fs the potential has reached its asymptotic value. As can be seen from the last plot, the difference between this potential and the Thomas-Fermi potential is rather small.

\begin{figure} 
\epsfxsize=14.0cm
\centerline{\epsffile{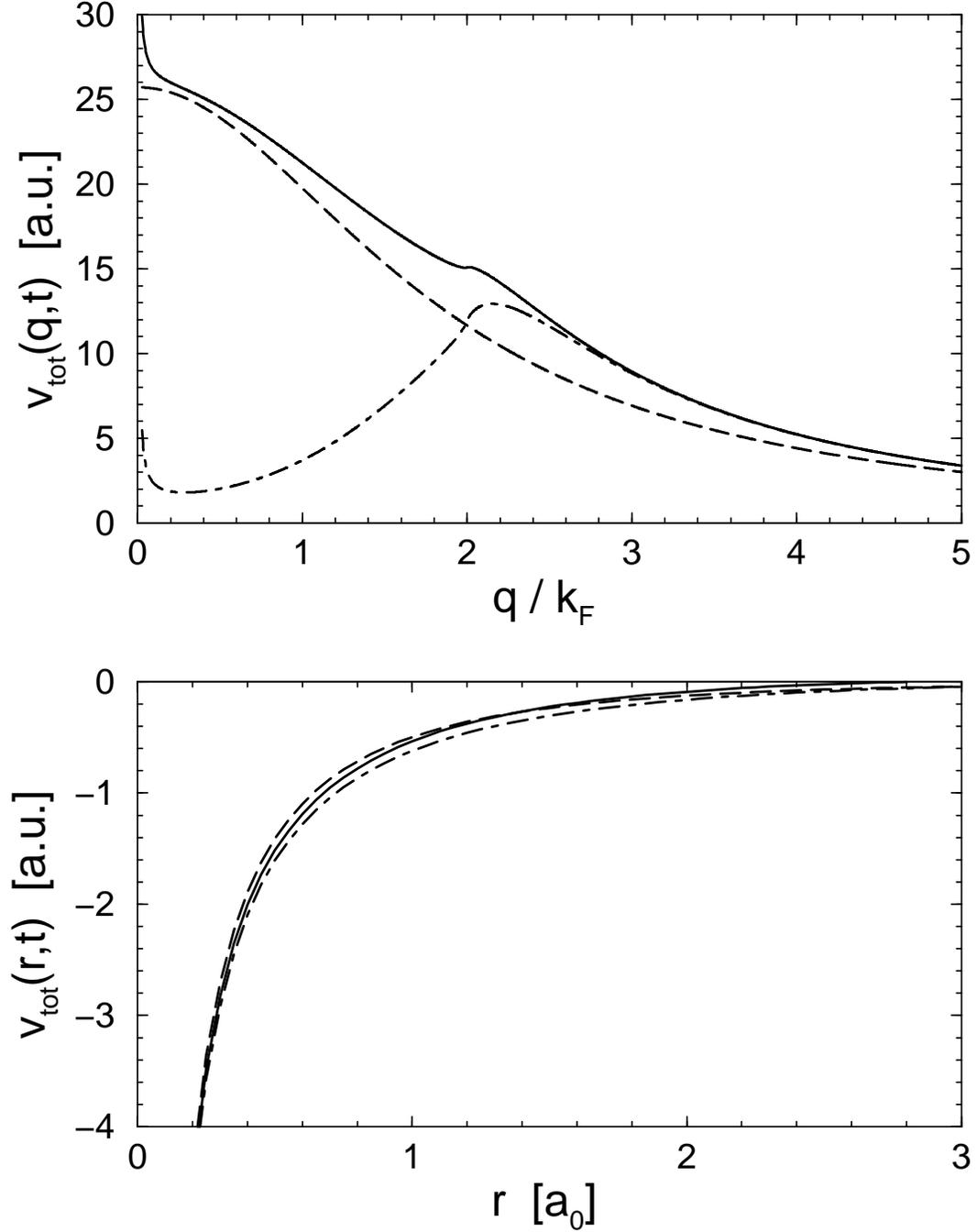}}
\caption{
Influence of the vertex correction $f\mib{xc}$ on the total potential. The upper panel shows 
$v\mib{tot}(q,t)$ calculated with and without the vertex correction (dot-dashed and solid line, 
respectively). In the lower panel $v\mib{tot}(r,t)$ is displayed. The solid line shows again the result of 
the calculation with $f\mib{xc}=0$ and the dot-dashed line the result of a calculation with a vertex 
correction. In both plots the dashed line denotes the Thomas-Fermi potential. The calculations where done for 
$r_s=5$ and $t=1000$ fs.
} 
\label{vJelfxc} 
\end{figure}

Before we close the discussion about the jellium model we want to discuss the influence of the vertex 
correction $f\mit{xc}$. We discuss this for $r_s=5$ since $f\mib{xc}$ for this Wigner-Seitz radius is more 
than eight times larger than the $f\mib{xc}$ for $r_s=2.07$ (-24.443 a.u. and -3.917 a.u., respectively). We 
use the parametrization of the exchange-correlation potential due to Perdew and Zunger. \cite{PeZu} The upper 
panel of Fig.~\ref{vJelfxc} shows a comparison of $v\mib{tot}(q,t)$ calculated with (dot-dashed line) and 
without (solid line) the use of vertex corrections, respectively. The figure shows that for wave vectors 
smaller than 2.2 $k_F$ the two curves start to differ. For small wave vectors the difference is quite 
remarkable. After the Fourier transform the difference in the resulting $v\mib{tot}(r,t)$ is much smaller. 
This is because the difference for small $q$ in $v\mib{tot}(q,t)$ is reflected in a difference for large $r$ 
in $v\mib{tot}(r,t)$. For our considerations this range is not of importance.


\section{Transient excitons in copper}
\label{copper}

\begin{figure} 
\epsfxsize=15.0cm
\centerline{\epsffile{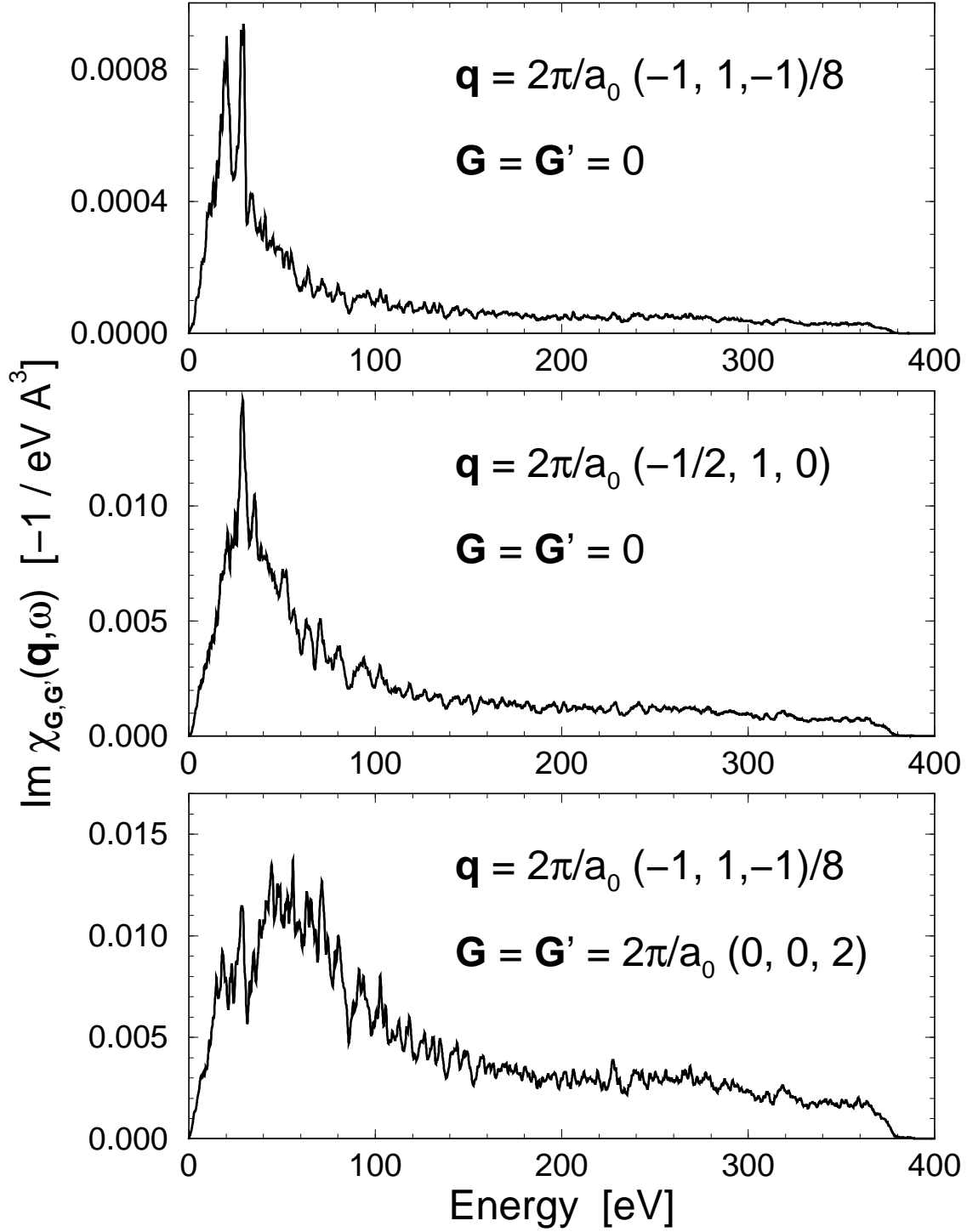}}
\caption{
The imaginary part of the density-response function $\chi_{\vecG=\vecG'}(\vecq,\omega)$ of Cu calculated for 
three wave vectors $\vecq+\vecG$. The underlying calculation of the polarizability was done including 200 
bands.
} 
\label{chiCu} 
\end{figure}

The heart of the calculation of the total potential Eq.~(\ref{vtotqt}) is the calculation of the 
polarizability given by Eq.~(\ref{pol}) for all wave vectors in the BZ and a sufficient number of $\vecG$ 
vectors needed to obtain a reliable density-response function Eq.~(\ref{chi}). In Fig.~\ref{chiCu} we show 
the imaginary part of $\chi_{\vecG,\vecG'}(\vecq.\omega)$ for $\vecG=\vecG'$ and three wave vectors 
$\vecq+\vecG$. The upper panel of Fig.~\ref{chiCu} shows the imaginary part of the density-response function 
for a small wave vector. It displays clearly the characteristic double peak structure \cite{igorCu} around 20 
and 29 eV sitting on top of a broad structureless hump which decays only very slowly. The calculations of the 
polarizability were done using 200 bands, i.e., considering transitions up to 355 -- 390 eV above the Fermi 
energy (the exact number depends on the wave vector). This energy cutoff is the reason for the sudden decline 
of ${\rm Im} \chi_{\vecG,\vecG'}(\vecq,\omega)$ between 360 and 380 eV which can be observed in all of the 
three plots. The middle panel shows ${\rm Im} \chi_{\vecG,\vecG'}(\vecq,\omega)$ for a wave vector on the 
edge of the BZ and the lower panel the imaginary part of the density-response function for a wave vector 
$\vecq+\vecG$ in the third zone. In all cases $\chi_{\vecG,\vecG'}(\vecq,\omega)$ was calculated within the 
RPA using a numerical broadening of $\eta=0.4$ eV in the calculation of the polarizability. It is obvious 
from the plots that this value is small compared to the broad structures of the density-response functions of 
Cu.  

In the jellium model the screening process was completely determined by the plasmon excitation. The plots in 
Fig.~\ref{chiCu} make clear that this is not the case for Cu. The density-response function of Cu is 
dominated by single-particle transitions and the already mentioned dispersionless double-peak structure caused by collective excitations of the {\it d\/} electrons. \cite{igorCu}

\begin{figure} 
\epsfxsize=15.0cm
\centerline{\epsffile{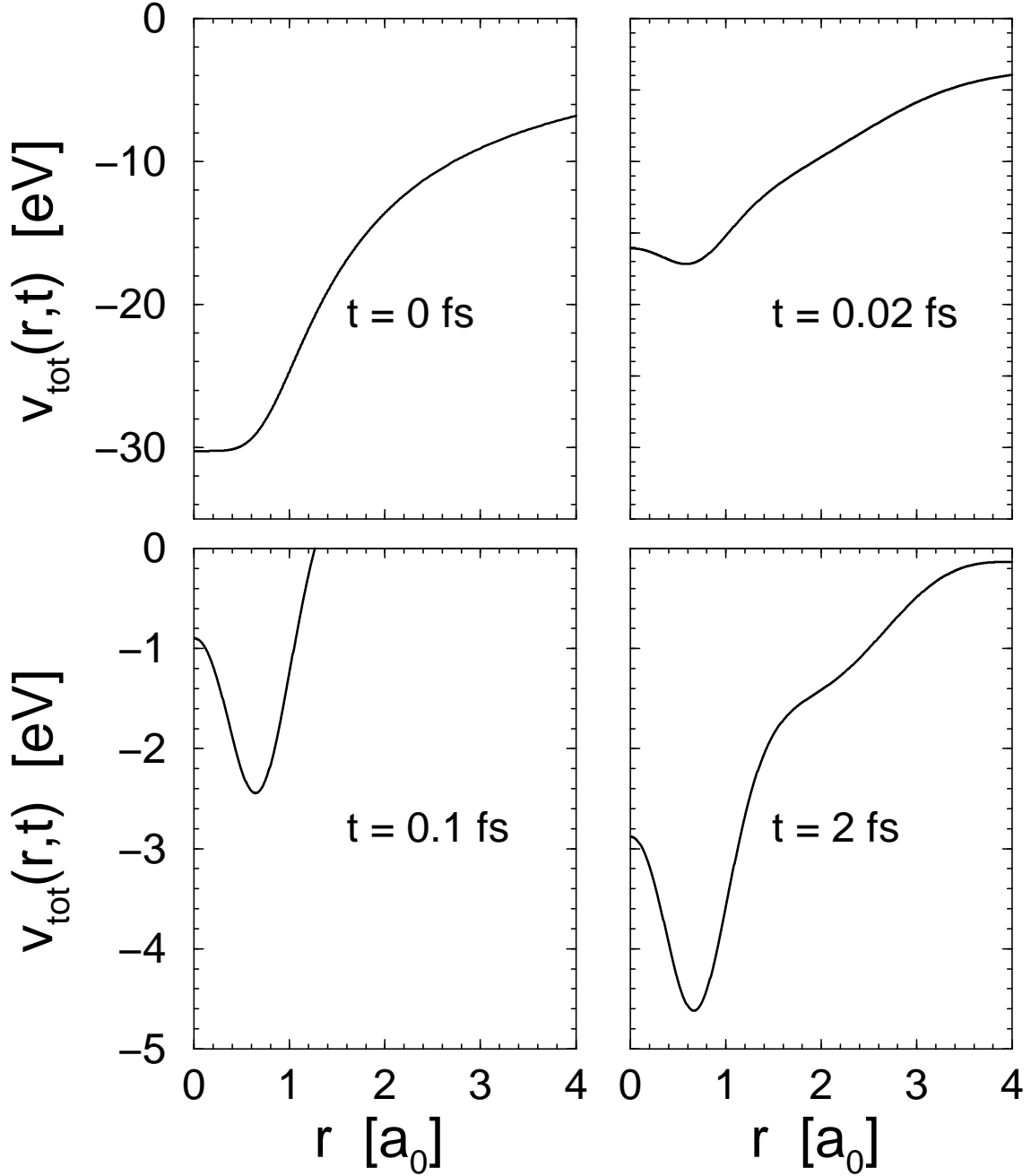}}
\caption{
Four snapshots of the total potential in Cu. The total potential was calculated using a polarizability which 
includes 200 bands. Note the different scales in the upper and lower row.
} 
\label{vCuSnaps} 
\end{figure}

As was already mentioned, the external potential Eq.~(\ref{vext}) is spherical symmetric. Although the full 
matrix of $\chi_{\vecG,\vecG'}(\vecq,\omega)$ was used in the calculation of the total potential 
$v\mib{tot}(\vecq+\vecG,t)$ (we used the first two shells of reciprocal lattice vectors, i.e., 15 $\vecG$ 
vectors) this leads to a nearly spherical symmetric total potential. Therefore the Fourier transform to real 
space could be done replacing the sum in Eq.~(\ref{vtotrt}) by an integral and performing the angular part of 
the integration analytically. Figure~\ref{vCuSnaps} shows the total potential in Cu as a function of $r$ for 
four different times. The plot in the upper left corner shows the potential at $t=0$ fs, i.e., the external 
potential. For finite times the potentials quickly becomes shallower. This behavior is the pendant to the 
time dependence within the jellium model where the potential became narrower.  As in jellium model the 
total potential is overscreening (see lower left panel in Fig.~\ref{vCuSnaps}) and oscillating around the 
equilibrium potential. However, compared to the jellium model this process is faster in Cu. The oscillations 
are completely died out at 2 fs. At this time $v\mib{tot}(r,t)$ has reached its asymptotic value. 

\begin{figure} 
\epsfxsize=14.0cm
\centerline{\epsffile{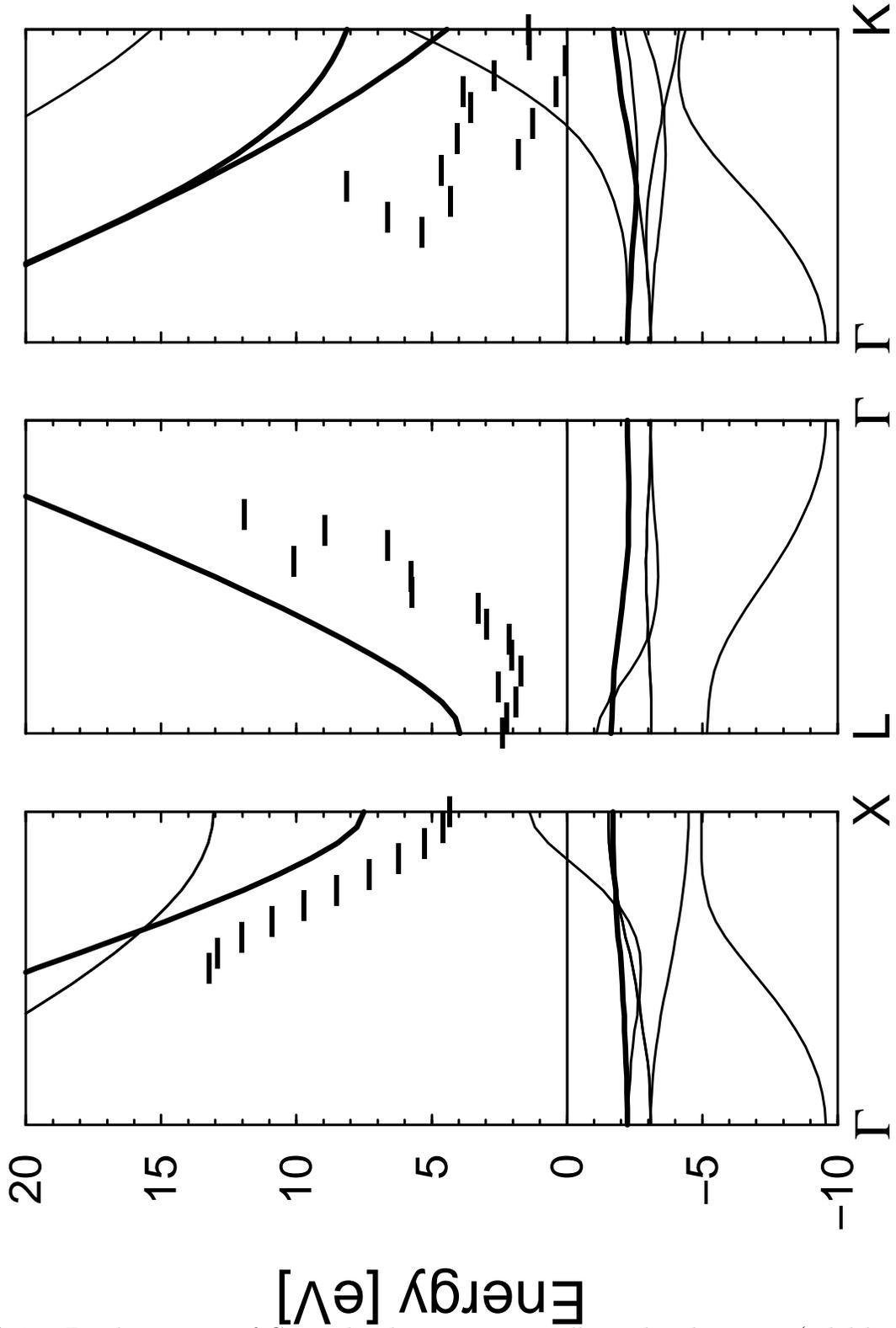}}
\caption{
Band structure of Cu in the three main crystallographic directions (solid lines). In addition to the  
single-particle states the transient excitonic states at $t=0.01$ fs ($L-\Gamma$ and $\Gamma-K$ directions) 
and $t=0.002$ fs ($\Gamma-X$ direction) are plotted as horizontal bars.
} 
\label{ExLev} 
\end{figure}

We are now in the position to apply the textbook approach for the estimate of excitonic binding energies in 
semiconductors to copper. We chose a high lying {\it d\/} band and (typically) the lowest unoccupied band for 
a given direction in the BZ and calculated the effective masses for each of the bands ($m_h^*$ and $m_c^*$ 
for the effective mass of the hole in the {\it d\/} band and the effective mass of the electron in the 
conduction band, respectively). A typical case is given by the two bands drawn as thick lines in the 
$L-\Gamma$ panel of Fig.~\ref{ExLev}. From the knowledge of $m_h^*$ and $m_c^*$ the reduced mass $\mu$ for 
each wave vector is determined. Then the Schr\"odinger equation for $v\mib{tot}(r,t)$ at a given time $t$ is 
solved for each $\mu$. $t$ has to be chosen in such a way that $v\mib{tot}(r,t)$ still possesses bound 
states. The resulting lowest eigenvalue is plotted with respect to the eigenvalue of the unoccupied band. 
This is the excitonic binding energy. 

In Fig.~\ref{ExLev} the excitonic binding energies in the three main crystallographic directions are drawn as
horizontal bars.  The bands which were used in the calculation are indicated by thick solid lines.  Since in
the $\Gamma-K$ direction there are two unoccupied bands close together, the calculation was done for both
bands.  Figure~\ref{ExLev} shows that excitonic states in the $L-\Gamma$ direction exist in an energy range
where there are no band states.  The energetic positions of the excitonic states are in agreement with the
experimental data (see Ref.~\onlinecite{PetekCu}).  The excitonic energy levels were determined from the
eigenvalues of a time-dependent potential at a fixed time.  They therefore shift with time.  However, in order
to study the effects of this shifting in more detail one needs to solve the Bethe-Salpeter equation
\cite{BSE1,BSE2,BSE3} which describes the dynamics of the correlated electron-hole pairs. This is beyond the scope of our model calculation which is intended to proof and explain the existence of transient excitonic states in Cu. In the
$\Gamma-K$ direction exists a {\it sp\/}-like band which crosses the Fermi energy.  In addition to these
single-particle states there are also a wealth of excitonic states.  Assuming that the lifetime of excitonic
states is longer than the electronic lifetime of band states (see Sec.~\ref{lifetime}) this explains the
experimental finding that the lifetime of electrons in these two directions are very similar.  It also
explains why $GW$ calculations -- which only address electron-electron scattering of excited band states with
valence states -- reveal shorter lifetimes as compared with experiment.  For both directions the eigenvalues
were calculated using the total potential $v\mib{tot}(r,t)$ at $t=0.01$ fs.  In the $\Gamma-X$ direction the
excitonic energies are a little bit higher due to the high energies of the lowest unoccupied band (7.5 eV at
the $X$ point).  The excitonic energies in this direction were calculated at $t=0.002$ fs.  Due to the large
slope of the lowest unoccupied bands the transient excitonic states which can be probed by experiment are all
in the vicinity of the high symmetry points $X$, $L$, and $K$.

 
\section{Lifetime of electrons in transient excitonic states}
\label{lifetime}

Having shown the possibility of transient excitons in metals we want to close with some comments about the 
lifetime of electrons in these states. The dominating process which determines the lifetime of excited 
electrons in band states is electron-electron scattering. Due to collisions with electrons of the valence 
band the excited electron is scattered out of its state, creating an electron-hole pair. This process -- 
which diagram is shown in Fig.~\ref{diagram}.a) -- and exchange scattering are the two processes which are 
considered in $GW$ calculations. However, only diagram \ref{diagram}.a) contributes to a final lifetime. 
Experience shows that the lifetime due to this process is proportional to the density of states of the final 
states. In metals this is therefore a rather efficient process. The bound states as described by transient 
excitons on the other hand can only decay via pair annihilation - pair creation, i.e., via an Auger process. The decay process is displayed in Fig.~\ref{diagram}.b). As a consequence of the bound-state character of electrons trapped in a transient excitonic state we conclude that the lifetime of these states is longer than in band states. This reasoning would explain the disagreement between {\it ab initio\/} $GW$ calculations for the lifetime of excited electrons along the $\Gamma-K$ direction and the experimental data. 
 

\section{conclusions}
\label{conclusions}

In this article we presented a model which allowed us to calculate transient excitonic states in metals. The 
model was applied to jellium and to real Cu. The results obtained within the jellium model describe how a 
(nearly) free electron system screens a localized positive charge. An analysis of the time-dependent total 
potential shows that the complete screening of the external charge takes 10 to 20 fs. This is the time needed 
until the polarizability is fully built up and in equilibrium. At this time the potential is nearly identical 
to the Thomas-Fermi potential. This near identity is caused by the fact that the screening process is completely determined by the plasmon excitations of the system. The contributions from single-particle transitions -- which are always present in the response function for finite wave vectors -- are negligible.  The potential possesses bound states for the first 0.2 fs. 

The situation in Cu is different since the imaginary part of the density-response function (which is
proportional to the loss function) has no distinct plasmon peak.  Instead the density-response function of Cu
shows plasmon-like excitations originating from the {\it d\/} electrons.  This double peak sits on top of a
broad structure caused by single-particle transitions.  The screening of a localized charge distribution is
therefore faster.  Nevertheless, the total potential possesses bound states for small times.  In our model
calculation for the determination of excitonic energies these eigenvalues are plotted with respect to the
lowest unoccupied band.  We find that the excitonic energies are in the energetic region of states detected in
the measurements of the lifetime of hot electrons in crystalline Cu using TR-2PPE spectroscopy.  It is
important to stress that only experiments with very short laser pulses allow the detection of these states.

Our calculations explain thus the puzzling experimental fact of states which are no band states. We interpret 
them as transient excitons. With the coarse estimate of the lifetime of these states we are also able to 
explain the difference between the experimentally determined lifetime of hot electrons in the (110) direction 
with respect to the results of $GW$ calculations; we think that the states probed in this direction are not 
band states but excitonic states. 

\begin{figure} 
\epsfxsize=5.0cm
\centerline{\epsffile{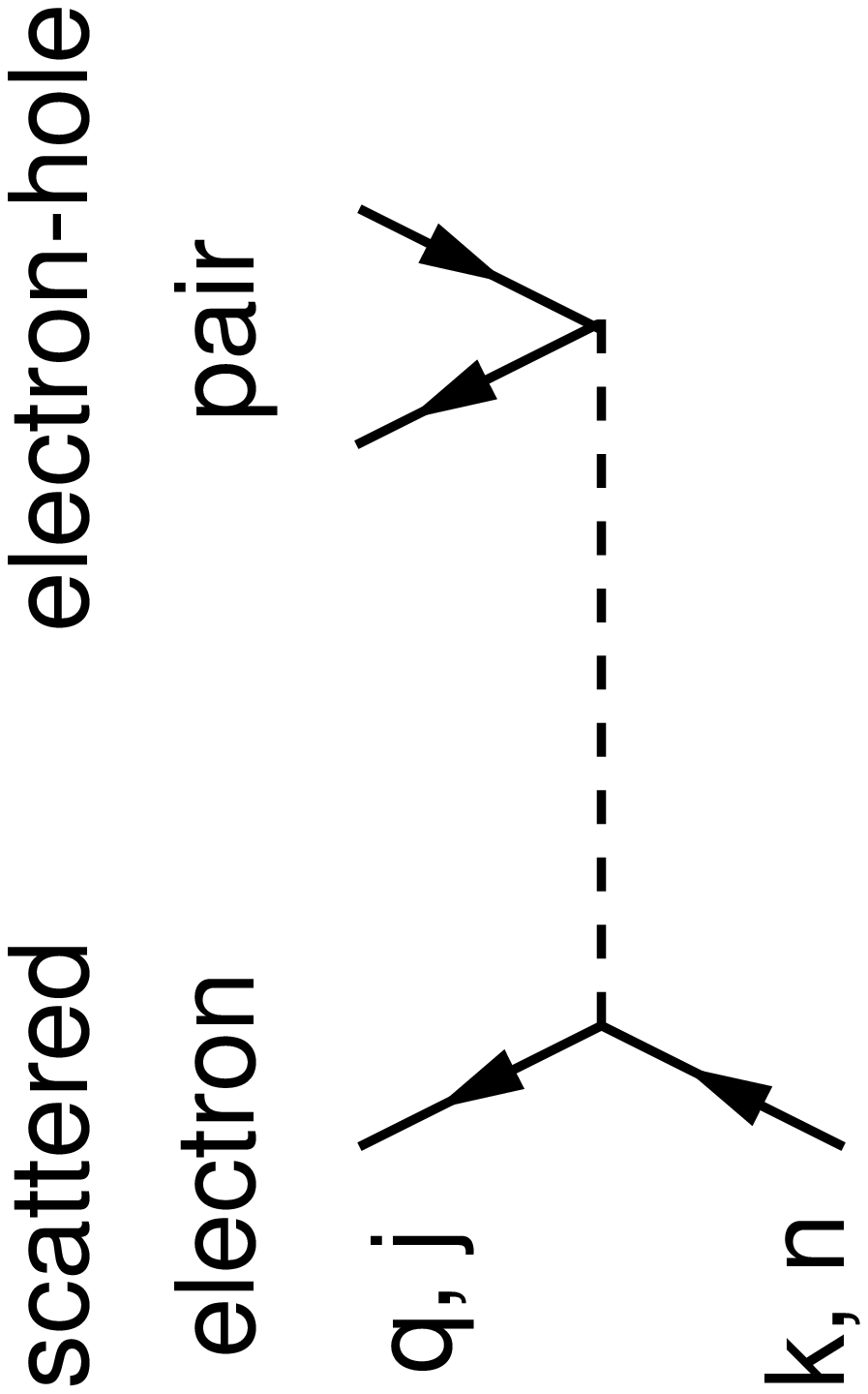}}
\epsfxsize=5.0cm
\centerline{\epsffile{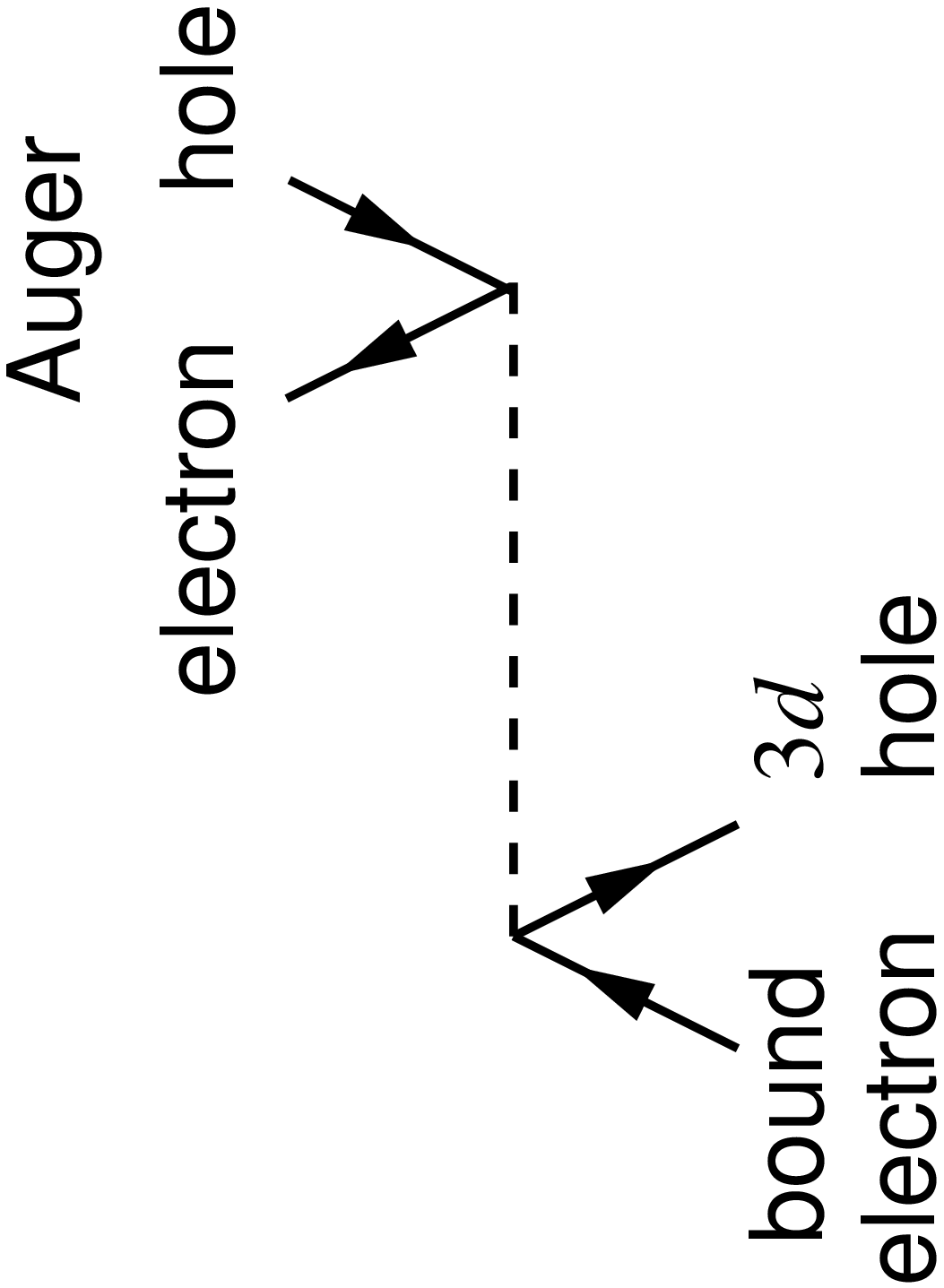}}
\caption{
Diagrams for the possible decay mechanism of electrons in a band state and a transient excitonic state. In a) 
the diagram for the decay of a band state is shown. This diagram determines the lifetime of excited electrons 
in $GW$ calculations. Diagram b) is the annihilation process involved in the decay of a transient excitonic 
state.
} 
\label{diagram} 
\end{figure}

Our model does not yet include the possibility to determine the exact lifetime of electrons in transient 
excitonic states. The model is only applicable up to the moment at which the electron is trapped in a bound 
state and an exciton is formed. Work along this line, i.e., the evaluation of diagram b) in 
Fig.~\ref{diagram} is in progress.

\acknowledgments 
G.  Ertl is gratefully acknowledged for his interest and generous support.  We thank R. Keyling for 
enlightening discussions.  This work is supported by the Deutsche Forschungsgemeinschaft through SFB 
450.



\end{document}